\documentstyle[a4,11pt,psfig]{article}
\begin{document}
%
%
%
%
\newcommand{\X}{\mbox{\boldmath X}}
\newcommand{\beq}{\begin{equation}}
\newcommand{\eeq}{\end{equation}}
\newcommand{\be}{\begin{equation}}
\newcommand{\ee}{\end{equation}}
\newcommand{\ie}{{\em i.e. }}
\newcommand{\al}{{\em et al }}
\newcommand{\eg}{{\em e.g. }}
\newcommand{\R}{{\it R }}
\newcommand{\ep}{{\epsilon }}
\newcommand{\eps}{{\epsilon }}
\newcommand{\nin}{\noindent}
\newcommand{\field}[1]{\mathbb{#1}}

\begin{center}
     {\Large
{\bf The Entropy of a Binary Hidden Markov Process}} \\

\vspace{4mm} \normalsize
Or Zuk$^1$, Ido Kanter$^2$ and Eytan Domany$^1$ \\
\vspace{4mm}

$^1$Dept. of Physics of Complex Systems, Weizmann Institute of
Science, Rehovot 76100, Israel\\
$^2$ Department of Physics, Bar Ilan University, Ramat Gan, 52900,
Israel
\end{center}

\vspace{1.7cm}


\begin{abstract}
{ \small  } \vspace{3mm} The entropy of a binary symmetric Hidden
Markov Process is calculated as an expansion in the noise
parameter $\epsilon$. We map the problem onto a one-dimensional
Ising model in a large field of random signs and calculate the
expansion coefficients up to second order in $\epsilon$. Using a
conjecture we extend the calculation to 11th order and discuss the
convergence of the resulting series.
\end{abstract}

\hspace{2mm} {\bf Key words: } Hidden Markov Process, entropy,
random-field Ising model

\newpage
\setlength{\topmargin}{-0.7in}
\section{Introduction }
%
\baselineskip 7mm {\it Hidden Markov Processes (HMPs)} have many
applications, in a wide range of disciplines - from the theory of
communication \cite{Merhav} to analysis of gene expression
\cite{Schliep}. Comprehensive reviews on both theory and
applications of {\it HMPs} can be found in (\cite{Merhav},
\cite{Rabiner}). Recent applications to experimental physics are
in (\cite{Kanter:01},\cite{Kanter:02}). The most widely used
context of {\it HMPs} is, however, that of construction of
reliable and efficient communication channels.

In a practical communication channel the aim is to reliably
transmit source message over a noisy channel. Fig 1. shows a
schematic representation of such a communication. The source
message can be a stream of words taken from a text. It is clear
that such a stream of words contains information, indicating that
words and letters are not chosen randomly. Rather, the probability
that a particular word (or letter) appears at a given point in the
stream depends on the words (letters) that were previously
transmitted. Such dependency of a transmitted symbol on the
precedent stream is modelled by a Markov process.

The Markov model is a finite state machine that changes state once
every time unit. The manner in which the state transitions occur
is probabilistic and is governed by a state-transition matrix,
$P$, that generates the new state of the system. The Markovian
assumption indicates that the state at any given time depends only
on the state at the previous time step. When dealing with text, a
state usually represents either a letter, a word or a finite
sequence of words, and the state-transition matrix represents the
probability that a given state is followed by another state.
Estimating the state-transition matrix is in the realm of
linguistics; it is done by measuring the probability of occurrence
of pairs of successive letters in a large corpus.

One should bear in mind that the Markovian assumption is very
restrictive and very few physical systems can expect to satisfy it
in a strict manner. Clearly, a Markov process imitates some
statistical properties of a given language, but can generate a
chain of letters that is grammatically erroneous and lack logical
meaning. Even though the Markovian description represents only
some limited subset of the correlations that govern a complex
process, it is the simplest natural starting point for analysis.
Thus one assumes that the original message, represented by a
sequence of $N$ binary bits, has been generated by some Markov
process. In the simplest scenario, of a binary symmetric Markov
process, the underlying Markov model is characterized by a single
parameter - the flipping rate $p$, denoting the probability that a
0 is followed by 1 (the same as a 1 followed by a 0). The stream
of $N$ bits is transmitted through a noisy communication channel.
The received string differs from the transmitted one due to the
noise. The simplest way to model the noise is known as the {\it
Binary Symmetric Channel}, where each bit of the original message
is flipped with probability $\epsilon$. Since the observer sees
only the received, noise-corrupted version of the message, and
neither the original message nor the value of $p$ that generated
it are known to him, what he records is the outcome of a {\it
Hidden Markov Process}. Thus, {\it HMPs} are double embedded
stochastic processes; the first is the Markov process that
generated the original message and the second, which does not
influence the Markov process, is the noise added to the Markov
chain after it has been generated.

Efficient information transmission plays a central role in modern
society, and takes a variety of forms, from telephone and
satellite communication to storing and retrieving information on
disk drives. Two central aspects of this technology are error
correction and compression. For both problem areas it is of
central importance to estimate $\Omega_R$, the number of
(expected) received signals.

In the noise free case this equals the expected number of
transmitted signals $\Omega_S$; when the Markov process has
flipping rate $p=0$, only two strings (all 1 or all 0) will be
generated and $\Omega_S = 2$, while when the flip rate is $p=1/2$
each string is equally likely and $\Omega_S=2^N$.

In general, $\Omega_R$ is given, for large $N$, by $2^{NH}$, where
$H=H(p,\epsilon)$ is the {\it entropy} of the process. The
importance of knowing $\Omega_R$ for compression is evident: one
can number the possible messages $i=1,2,...\Omega_R$, and if
$\Omega_R < 2^N$, by transmitting only the index of the message
(which can be represented by $\log _2 \Omega_R  < N$ bits) we
compress the information. Note that we can get further compression
using the fact that the $\Omega_R$ messages do not have equal
probabilities.

Error correcting codes are commonly used in methods of information
transmission to compensate for noise corruption of the data during
transmission. These methods require the use of additional
transmitted information, i.e., redundancy, together with the data
itself. That is, one transmits a string of $M>N$ bits; the
percentage of additional transmitted bits required to recover the
source message determines the coding efficiency, or channel
capacity, a concept introduced and formulated by Shannon. The
channel capacity for the BSC and for a random i.i.d. source was
explicitly derived by Shannon in his seminal paper of 1948
\cite{Shannon}. The calculation of channel capacity for a
Markovian source transmitted over a noisy channel is still an open
question.

Hence, calculating the entropy  of a {\it HMP} is an important
ingredient of progress towards deriving improved estimates of both
compression and channel capacity, of  both theoretical and
practical importance for modern communication. In this paper we
calculate the entropy of a {\it HMP} as a power series in the
noise parameter $\epsilon$.

In Section 2 we map the problem onto that of a one-dimensional
nearest neighbor Ising model in a field of fixed magnitude and
random signs (see \cite{Nattermann} for a review on the Random
Field Ising Model). Expansion in $\epsilon$ corresponds to working
near the infinite field limit.

Note that the object we are calculating is {\it not} the entropy
of an Ising chain in a quenched random field, as shown in eq.
(\ref{H_N_def}) and in the discussion following it. In technical
terms, here we set the replica index to $n=1$ after the
calculation, whereas to obtain the (quenched average) properties
of an Ising chain one works in the $n \rightarrow 0$ limit.

In Sec. 3 we present exact results for the expansion coefficients
of the entropy up to second order. While the zeroth and first
order terms were previously known (\cite{Jacquet},
\cite{Weissman:02}), the second order term was not \cite{ZKD1} .
In Sec 4. we introduce bounds on the entropy that were derived by
Cover and Thomas \cite{Cover}; we have strong evidence that these
bounds actually provide the exact expansion coefficients. Since we
have not proved this statement, it is presented as a conjecture;
on it's basis the expansion coefficients up to eleventh order are
derived and listed. We conclude in Sec. 5 by studying the radius
of convergence of the low-noise expansion, and summarize our
results in Sec 6.
\section{A Hidden Markov Process and the Random-Field Ising Model}
\subsection{Defining the process and its entropy}
Consider the case of a binary signal generated by the source.
Binary valued symbols, $s_i=\pm 1$ are generated and transmitted
at fixed times $i \Delta t $. Denote a sequence of $N$ transmitted
symbols by
 \beq S = (s_1,s_2,....s_N) \eeq
The sequence is generated by a Markov process; here we assume that
the value of $s_{i+1}$ depends only on $s_i$ (and not on the
symbols generated at previous times). The process is parametrized
by a transition matrix $P$, whose elements are the transition
probabilities 
\beq P_{+,-}= Pr(s_{i+1}=+1|s_i=-1) \qquad \qquad P_{-,+}=
Pr(s_{i+1}=-1|s_i=+1) \eeq
Here we treat the case of a {\it symmetric} process, i.e. $P_{+,-}
= P_{-,+}=p,~~$ so that we have
\beq s_{i+1} = \left\{  \begin{array}{ll}
                      ~~ s_i \qquad \qquad & {\rm prob.}= 1-p \\
                     -s_i \qquad \qquad & {\rm prob.}= p \end{array} \right.
\label{eq:Markov} \eeq
The first symbol $s_1$ takes the values $\pm 1$ with equal
probabilities, $Pr(s_1=+1)=Pr(s_1=-1)=1/2$. The probability of
realizing a particular sequence $S$ is given by
\beq Pr(S) = \frac{1}{2}\prod_{i=2}^N Pr(s_i|s_{i-1})
\label{eq:P(S)} \eeq
The generated sequence $S$ is "{\it sent}" and passes through a
noisy channel; hence the {\it received} sequence,
\beq \R = (r_1,r_2,...r_N) \eeq
is not identical to the transmitted one. The noise can flip a
transmitted symbol with probability $\epsilon$:
\beq Pr(r_i=-s_i|s_i)=\epsilon \label{eq:f}\eeq
Here we assumed that the noise is generated by an independent
identically distributed (iid) process; the probability of a flip
at time $i$ is independent of what happened at other times $j<i$
and of the value of $i$. We also assume that the noise is
symmetric, i.e. the flip probability does not depend on $s_i$.

Once the underlying Markov process $S$ has been generated, the
probability of observing a particular sequence $\R$ is given by
\beq Pr(\R|S) = \prod_{i=1}^N Pr(r_i|s_i) \label{eq:PSig} \eeq
and the joint probability of any particular $S$ and $\R$ to occur
is given by
\beq Pr(\R, S) = Pr(\R|S)Pr(S) \label{eq:joint} \eeq
The original transmitted signal, $S$, is "hidden" and only the
received (and typically corrupted) signal $\R$ is "seen" by the
observer. Hence it is meaningful to ask - what is the probability
to observe any particular received signal $\R$? The answer is 
\beq  Q(\R ) = \sum_S Pr(\R , S) \label{eq:QSig} \eeq
Furthermore, one is interested in the\footnote{The Shannon entropy
is defined using log$_2$; we use natural log for simplicity} {\it
Shannon entropy H} of the observed process,
\beq H_N = -\sum_\R Q(\R) \log Q(\R) \label{eq:HN} \eeq
and in particular, in the {\it entropy rate}, defined as
\beq H = \lim_{N \rightarrow \infty} \frac{H_N}{N}\eeq
\subsection{Casting the problem in Ising form}
It is straightforward to cast the calculation of the entropy rate
onto the form of a one-dimensional Ising model. The conditional
Markov probabilities (\ref{eq:Markov}), that connect the symbols
from one site to the next, can be rewritten as
\beq Pr(s_{i+1}|s_i) = e^{J s_{i+1}s_i}/(e^J + e^{-J}) \qquad {\rm
with} \qquad e^{2J} = (1-p)/p \label{eq:pJ} \eeq 
and similarly, the flip probability generated by the noise,
(\ref{eq:f}), is also recapitulated by the Ising form
\beq Pr(r_i|s_i)= e^{K r_i s_i}/(e^K+e^{-K}) \qquad {\rm with}
\qquad e^{2K}=(1-\epsilon)/\epsilon \eeq
The joint probability of realizing a pair of transmitted and
observed sequences ($S, \R$) takes the form
(\cite{Saul},\cite{MacKay})
\beq Pr(\R,S)= A \exp \left( J \sum_{i=1}^{N-1} s_{i+1} s_i + K
\sum_{i=1}^N r_i s_i \right) \label{eq:prIsing} \eeq
where the constant $A$ is the product of two factors, $A=A_0 A_1$,
given by
\beq A_0 = \frac{1}{2} \left( e^J + e^{-J} \right)^{-(N-1)} \qquad
\qquad A_1 = \left( e^K + e^{-K} \right)^{-N}  \eeq
The first sum in (\ref{eq:prIsing}) is the Hamiltonian of a chain
of Ising spins with open boundary conditions and nearest neighbor
interactions $J$; the interactions are ferromagnetic ($J>0$) for
$p < 1/2$. The second term corresponds, for small noise $\epsilon
<1/2$, to a strong ferromagnetic interaction $K$ between each spin
$s_i$ and another spin, $r_i$, connected to $s_i$ by a "dangling
bond" (see Fig 2).

Denote the summation over the hidden variables by $Z(\R)$:
\beq Z(\R) = \sum_S \exp \left( J \sum_{i=1}^{N-1} s_{i+1} s_i + K
\sum_{i=1}^N r_i s_i \right)  \label{eq:ZR} \eeq
so that the probability $Q(\R)$ becomes (see eq. (\ref{eq:QSig}))
$Q(\R ) = A~Z(\R)$. Substituting in (\ref{eq:HN}),  the entropy of
the process can be written as
\beq
H_N =  - \sum_\R A~Z(\R) \log[A Z(\R)]
    =  - \left[ \frac{d }{dn} \sum_\R [A~Z(\R)]^n
    \right]_{n=1}
\label{H_N_def} \eeq

The interpretation of this expression is obvious: an Ising chain
is submitted to local fields $h_i=Kr_i$, with the sign of the
field at each site being $\pm$ with equal probabilities, and we
have to average $Z(h_1,...h_N)^n$ over the field configurations.
This is precisely the problem one faces in order to calculate
properties of a well-studied model, of a nearest neighbor Ising
chain in a quenched random field of uniform strength and random
signs at the different sites (there one is interested, however, in
the limit $n \rightarrow 0$). This problem has not been solved
analytically, albeit a few exactly solvable simplified versions of
the model do exist (\cite{Derrida}, \cite{Fisher},
\cite{Grinstein}, \cite{Nieuwenhuizen}), as well as expansions
(albeit in the weak field limit \cite{DerHil}).

One should note that here we calculate the entropy associated with
the observed variables \R. In the Ising language this corresponds
to an entropy associated with the randomly assigned signs of the
local fields, and {\it not} to the entropy of the spins $S$.
Because of this distinction the entropy $H_N$ has no obvious
physical interpretation or relevance, which explains why the
problem has not been addressed yet by the physics community.

We are interested in calculating the entropy rate {\it in the
limit of small noise}, i.e. $\epsilon \ll 1$. In the Ising
representation this limit corresponds to $K \gg 1$ and hence an
expansion in $\epsilon$ corresponds to expanding near the infinite
field limit of the Ising chain.

\section{Expansion to order $\epsilon^2$: exact results}
We are interested in calculating the entropy rate
\beq H = - \lim_{N \rightarrow \infty}\left[ \frac{1}{N}\sum_\R
A~Z(\R) \log A~Z(\R)\right] \label{eq:HI} \eeq
to a given order in $\epsilon$. A few technical points are in
order. First, we will actually use
\be  e^{-2K}=\epsilon/(1-\epsilon) \ee as our small parameter and
expand to order $\epsilon^2$ afterwards. Second, we will calculate
$H_N$ and take the large $N$ limit. Therefore we can replace the
open boundary conditions with periodic ones (setting
$s_{N+1}=s_1$) - the difference is a surface effect of order
$1/N$. The constant $A_0$ becomes \be A_0 =\left( e^J + e^{-J}
\right)^{-N} \ee and the interaction term $J s_1s_N$ is added to
the first sum in eq. (\ref{eq:prIsing}), which contains now $N$
pairs of neighbors.

{\bf Expanding $Z(\R)$:} Consider $Z(\R )$ from (\ref{eq:ZR}). For
any fixed \R$=(r_1,r_2,...r_N)$ the leading order is obtained by
the $S$ configuration with $s_i=r_i$ for all $i$. For this
configuration each site contributes $K$ to the "field term" in
(\ref{eq:ZR}). The contribution of this configuration to the
summation over $S$ in (\ref{eq:ZR}) is
\beq Z(\R)^{(0)} = e^{NK} \exp \left( J \sum_{i=1}^N r_{i+1} r_i
\right) \label{eq:Z0} \eeq
The next term we add consists of the contributions of those $S$
configurations which have $s_i=r_i$ at {\it all but one position}.
The field term of such a configuration is $K$ from $N-1$ sites and
$-K$ from the single site with $s_j=-r_j$. There are $N$ such
configurations, and the total contribution of these terms to the
sum (\ref{eq:ZR}) is
\beq Z(\R)^{(1)} = e^{NK}e^{-2K} \exp \left( J \sum_{i=1}^N
r_{i+1} r_i \right) \sum_{j=1}^N \exp [-2J r_j(r_{j-1}+r_{j+1})]
\label{eq:Z1} \eeq 
The next term is of the highest order studied in this paper; it
involves configurations $S$ with all but two spins in the state
$s_i=r_i$; the other two take the values $s_j=-r_j,~~ s_k=-r_k$,
i.e. are flipped with respect to the corresponding local fields.
These $S$ configurations belong to one of two classes. In class
$a$ the two flipped spins are located on nearest neighbor sites,
e.g. $k=j+1$; there are $N$ such configurations. To the second
class, $b$, belong those configurations in which the two flipped
spins are {\it not} neighbors - there are $N(N-3)/2$ such terms in
the sum (\ref{eq:ZR}), and the respective contributions
are\footnote{ We use the obvious identifications imposed by
periodic boundary conditions, e.g. $r_{N+1}=r_1,~~r_{N+2}=r_2$} 
\beq Z(\R)^{(2a)} = e^{NK}e^{-4K} \exp \left( J \sum_{i=1}^N
r_{i+1} r_i \right) \sum_{j=1}^N \exp [-2J
(r_jr_{j-1}+r_{j+1}r_{j+2})] \label{eq:Z2a} \eeq 
\beq Z(\R)^{(2b)} = e^{NK}e^{-4K} \exp \left( J \sum_{i=1}^N
r_{i+1} r_i \right) \frac{1}{2}\sum_{j=1}^N \sum_{k \neq j,j\pm1}
\exp [-2J r_j(r_{j-1}+r_{j+1}) - 2J r_k(r_{k-1}+r_{k+1})]
\label{eq:Z2b} \eeq
Calculation of $H$ is now straightforward, albeit tedious:
substitute $AZ$ into eq. (\ref{eq:HI}), expand everything in
powers of $\epsilon$, to second order, and for each term perform
the summation over all the $r_i$ variables. These summations
involve two kinds of terms. The first is of the
"partition-sum-like" form
\beq \sum_R e^{{\cal H}(R)} \qquad  {\rm where} \qquad {\cal
H}(R)=\sum_j \Delta_j J r_j r_{j+1} \qquad {\rm with} \qquad
\Delta_j=\pm 1 \eeq
For the case studied here we encounter either all bonds $\Delta_j
J>0$, or two have a flipped sign (corresponding to eq.
(\ref{eq:Z1}, \ref{eq:Z2a})), or four have flipped signs
(corresponding to (\ref{eq:Z2b})). These "partition-sum-like"
terms are independent of the signs of the $\Delta_j$; in fact we
have for all of them
\beq  A_0 \sum_R e^{{\cal H}(R)} =1 \eeq
The second type of term that contributes to $H$ is of the
"energy-like" form:
\beq \sum_R e^{{\cal H}(R)} r_k r_{k+1}  \eeq
The absolute value of these terms is again independent of the
$\Delta_j$, but one has to keep track of their signs. Finally, one
has to remember that the constant $A_1$ also has to be expanded in
$\epsilon$. The calculation finally yields the following result
(here we switch from $J$ to the "natural" variable $p$ using eq.
(\ref{eq:pJ})):
\beq H(p,\epsilon) = \sum_{k=0}^{\infty} H^{(k)}(p) \epsilon^k
\label{eq:exp} \eeq with the coefficients $H_k$ given by :
\beq H^{(0)} = -p \log p - (1-p) \log(1-p) \qquad \qquad H^{(1)} =
2 (1-2p) \log \left[ \frac{1-p}{p} \right] \label{eq:H1H2} \eeq
\beq H^{(2)} = -2(1-2p) \log \left[ \frac{1-p}{p} \right]
-\frac{(1-2p)^2}{2p^2(1-p)^2} \label{012_stat}\eeq The zeroth and
first order terms (\ref{eq:H1H2}) were known (\cite{Jacquet},
\cite{Weissman:02}), while the second order term is new
\cite{ZKD1}.

\section{Upper Bounds derived using a system of finite length}
\nin When investigating the limit $H$, it is useful to study the
quantity $C_N = H_N - H_{N-1}$, which is also known as the {\it
conditional} entropy. $C_N$ can be interpreted as the average
amount of uncertainty we have on $r_N$, assuming that we know
$(r_1,\ldots,r_{N-1})$. Provided that $H$ exist, it easily follows
that \be H = \lim_{N \to \infty} C_N \ee

\nin Moreover, according to \cite{Cover}, $C_N \geq H$, and the
convergence is monotone : \be C_N \searrow H \quad (N \to \infty)
\ee

\nin We can express $C_N$ as a function of $p$ and $\epsilon$ by
using eq. (\ref{H_N_def}). For this, we represent $Z$ using the
original variables $p, \eps$ (note that from this point of we work
with open boundary conditions on the Ising chain of $N$ spins):

\be Z(R) = \sum_S (1-p)^{\sum_{i=1}^{N-1} 1_{S_i = S_{i+1}}}
p^{N-1-\sum_{i=1}^{N-1} 1_{S_i = S_{i+1}}} (1-\eps)^{\sum_{i=1}^N
1_{S_i = R_i}}  \eps^{N-\sum_{i=1}^n 1_{S_i = R_i}}
\label{prob_R_cs_style} \ee
 \nin where we denote $1_{s,s'}=(1+s s')/2$. Eq.
 (\ref{prob_R_cs_style}) gives $Z(R)$ explicitly as a polynomial in $p$ and $\eps$ with
maximal degree $N$, and can be represented as :

\be Z(R) = \sum_{i=0}^N Z_{i} (R) \eps^i \label{Q_def}\ee

\nin Here $Z_{i} = Z_{i}(R)$ are functions of $p$ only.

\nin Substituting this expansion in eq. (\ref{H_N_def}), and
expanding $\log Z(R)$ according to the Taylor series $\log(a+x) =
\log(a) - \sum_{n=1}^{\infty} \frac{(-x)^n}{n a^n}$, we get

\be H_N = -\sum_{R} \left[ \sum_{i=0}^N Z_{i} (R) \eps^i \right]
\Biggl[ \log Z_0(R) - \Biggr. \left. \sum_{j=1}^k \frac{
(-\sum_{i=1}^n Z_{i} (R) \eps^i)^j }{j Z_0 (R)^j} \right] +
O(\eps^{k+1}) \ee

\nin When extended to terms of order $\epsilon^k$, this equation
gives us precisely the expansion of the upper-bound $C_N$ up to the
{\it k-th} order,
 \beq
 C_N = \sum_{i=0}^{k} C_N^{(i)}\eps^i + O(\eps^{k+1})
  \eeq
 For example, stopping at order $L=2$ gives
$$
H_N = - \sum_{Y} \biggl\{ Z_0(R) \log Z_0(R) + \biggr. \left[
Z_1(R) (1 + \log Z_0(R)) \right] \eps +
$$
\be \left. \left[ \frac{Z_1(R)^2}{2Z_0(R)} + Z_2(R) (1 + \log
Z_0(R)) \right] \eps^2 \right\}+ O(\eps^3)
\label{entropy_expansion} \ee

\nin The zeroth and first order terms can be evaluated
analytically for any $N$; beyond first order, we can compute the
expansion of $H_N$ symbolically \footnote{The computation we have
done is exponential in $N$, but the complexity can be improved. }
(using Maple \cite{Maple}), for any finite $N$. This was actually
done, for $N \leq 8$ and $k \leq 11$. For the first order we have
proved (\cite{ZKD1}) that $C_N^{(1)}$ is independent of $N$ (and
equals $H^{(1)}$). The symbolic computation of higher order terms
yielded similar independence of $N$, provided that $N$ is large
enough. So, $C_N^{(k)} = C^{(k)}$ for large enough $N$. For
example, $C_N^{(2)}$ is independent of $N$ for $3 \leq N \leq 8$
and equals the exact value of $H^{(2)}$ as given by eq.
(\ref{012_stat}). Similarly, $C_N^{(4)}$ settles, for $N \geq 4$,
at some value denote by $C^{(4)}$, and so on. For the values we
have checked, the settling point for $C_N^{(k)}$ turned out to be
at $N=\lceil \frac{k+3}{2} \rceil$. This behavior is, however,
unproved for $k \geq 2$, and, therefore, we refer to it as a

{\bf Conjecture:} For any order $k$, there is a critical chain
length $N_c(k)=\lceil \frac{k+3}{2} \rceil$ such that for
$N>N_c(k)$ we have $C_N^{(k)} = C^{(k)}$. \\

It is known that $C_N \to H$, and $C_N$ and $H$ are analytic
functions of $\eps$ at $\eps=0$ \footnote{See next section on the
Radius of Convergence}, so that we can expand both sides around
$\eps = 0$, and conclude that $C_N^{(k)} \to H^{(k)}$ for any $k
\geq 1$ when $N \to \infty$. Therefore, if our conjecture is true,
and $C_N^{(k)}$ indeed settles  at some value $C^{(k)}$
independent of $N$ (for $N>N_c(k)$ ), it immediately follows that
this value equals $H^{(k)}$. Note that the settling is rigourously
supported for $k=0,1$, while for $k=2$ we showed that indeed
$C^{(2)}=H^{(2)}$, supporting our conjecture.

\nin The first orders up to $H^{(11)}$, obtained by identifying
$H^{(k)}$ with $C^{(k)}$, are given in the Appendix, as functions
of $\lambda= 1-2p$, for better readability. The values of
$H^{(0)},H^{(1)}$ and $H^{(2)}$ coincide with the results that
were derived rigorously from the low-temperature/high-field
expansion, thus giving us support for postulating the above
Conjecture.

\nin Interestingly, the nominators have a simpler expression when
considered as a functions of $\lambda$, which is the second
eigenvalue of the Markov transition matrix $P$. Note that only even
powers of $\lambda$ appear. Another interesting observation is that
the free element  in $[p(1-p)]^{2(k-1)} H^{(k)}$ (when treated as a
polynomial in $p$), is $\frac{(-1)^k}{k(k-1)}$, which might suggest
some role for the function $\log(1+\frac{\eps}{[2p(1-p)]^2})$ in the
first derivative of $H$. All of the above observations led us to
conjecture the following form for $H^{(k)}$ (for $k \geq 3$) : \be
H^{(k)} = \frac{2^{4(k-1)} \sum_{j=0}^{d_k} a_{j,k} \lambda^{2j} }{
k(k-1) (1-\lambda^2)^{2(k-1)}} \ee
 where $a_{j,k}$ and $d_k$ are integers that can be seen in the
 Appendix for $H^{(k)}$ up to $k=11$.

\section{The Radius of Convergence}
If one wants to use our expansion around $\eps=0$ for actually
estimating $H$ at some value $\eps$, it is important to ascertain
that $\eps$ lies within the radius of convergence of the
expansion. The fundamental observation made here is that for
$p=0$, the function $H(\eps)$ is not an analytic function at $\eps
= 0$, since its first derivative diverges. As we increase $p$, the
singularity points 'moves' to negative values of $\eps$, and hence
the function is analytic at $\eps = 0$, but the radius of
convergence is determined by the distance of $\eps = 0$ from this
singularity. Denote by $\rho(p)$ the radius of convergence of
$H(\eps)$ for a given $p$;  we expect that $\rho(p)$ grows when we
increase $p$, while for $p \to 0$, $\rho(p) \to 0$.

It is useful to first look at a simpler model, in which there is
no interaction between the spins. Instead, each spin is in an
external field which has a uniform constant component $J$, and a
site-dependent component of absolute value $K$ and a random sign.
For this simple i.i.d. model the entropy rate takes the form
 \be H
= h_b[p(1-\eps)+ \eps(1-p)] \label{entropy_iid}\ee
 \nin where $h_b[x] =
-[x \log x + (1-x) \log (1-x)]$ is the binary entropy function.
Note that for $\eps =0$ the entropy of this model equals that of
the Ising chain. Expanding eq. (\ref{entropy_iid}) in $\eps$ (for
$p
> 0$) gives :
$$
H = -(p \log p + (1-p) \log (1-p)) + (1-2p) \log (\frac{1-p}{p})
\eps +
$$
\be
\sum_{k=2}^{\infty} \frac{1}{k(k-1)} \left[ \frac{(2p-1)^k}{p^k} +
\frac{(1-2p)^k}{(1-p)^{k-1}} \right] \eps^k \ee
 \nin The radius of convergence here is easily shown to be
$p/(1-2p)$; it goes to 0 for $p \rightarrow 0$ and increases
monotonically with $p$.

\nin Returning to the {\it HMP}, the orders $H^{(k)}$ are (in
absolute value) usually larger than those of the simpler i.i.d.
model, and hence the radius of convergence may be expected to be
smaller. Since we could not derive $\rho(p)$ analytically, we
estimated it using extrapolation based on the first $11$ orders.
We use the fact that $\rho(p) = \lim_{k \to \infty}
\frac{H^{(k)}}{H^{(k+1)}}$ (provided the limit exists). The data
was fitted to a rational function of the following form (which
holds for the i.i.d. model):

\be \frac{H^{(k)}}{H^{(k+1)}} \sim \frac{a k + b}{k + c}, \ee \nin
For a given fit, the radius of convergence was simply estimated by
$a$. The resulting prediction is given in Fig.
\ref{ConvergenceRadius_fig} for both the i.i.d. model (for which
it is compared to the known exact $\rho(p)$) and for the {\it
HMP}. While quantitatively, the predicted radius of the {\it HMP}
is much smaller than this of the i.i.d. model, it has the same
qualitative behavior, of starting at zero for $p=0$, and
increasing with $p$.

\nin We compared the analytic expansion to estimates of the
entropy rate based on the lower and upper bounds, for two values
of $\eps$ (see Fig. \ref{OrdersApproximationTwoEps_fig})  . First
we took $\eps = 0.01$, which is realistic in typical communication
applications. For $p$ less than about 0.1 this value of $\eps$
exceeds the radius of convergence and the series expansion
diverges, whereas for larger $p$ the series converges and gives a
very good approximation to $H(p, \eps=0.01)$. The second value
used was $\eps = 0.2$; here the divergence happens for $p \leq
0.37$, so the expansion yields a good approximation for a much
smaller range. We note that, as expected, the approximation is
much closer to the upper bound than to the lower bound of
\cite{Cover}.

\section{Summary}
Transmission of a binary message through a noisy channel is
modelled by a Hidden Markov Process. We mapped the binary
symmetric {\it HMP} onto an Ising chain in a random external field
in thermal equilibrium. Using a low-temperature/high-random-field
expansion we calculated the entropy of the {\it HMP} to second
order $k=2$ in the noise parameter $\epsilon$. We have shown for
$k \leq 11$ that when the known upper bound on the entropy rate is
expanded in $\eps$, using finite chains of length $N$, the
expansion coefficients settle, for $N_c(k) \leq N \leq 8$, to
values that are independent of $N$. Posing a conjecture, that this
continues to hold for any $N$, we identified the expansion
coefficients of the entropy up to order 11. The radius of
convergence of the resulting series was studied and the expansion
was compared to the the known upper and lower bounds.

By using methods of Statistical Physics we were able to address a
problem of considerable current interest in the problem area of
noisy communication channels and data compression.

\section*{Acknowledgments}
I.K. thanks N. Merhav for very helpful comments, and the Einstein
Center for Theoretical Physics for partial support. This work was
partially supported by grants from the Minerva Foundation and by
the European Community's Human Potential Programme under contract
HPRN-CT-2002-00319, STIPCO.

\section*{Appendix}
\nin Orders three to eleven, as function of $\lambda = 1-2p$.
(Orders $0-2$ are given in equations (\ref{eq:H1H2} -
\ref{012_stat})) :


$$
H^{(3)} =
\frac{-16(5\lambda^4-10\lambda^2-3)\lambda^2}{3(1-\lambda^2)^4}
$$

$$
H^{(4)} =
\frac{8(109\lambda^8+20\lambda^6-114\lambda^4-140\lambda^2-3)\lambda^2}{3(1-\lambda^2)^6}
$$

$$
H^{(5)} =
\frac{-128(95\lambda^{10}+336\lambda^8+762\lambda^6-708\lambda^4-769\lambda^2-100)\lambda^4}{15(1-\lambda^2)^8}
$$

$$
H^{(6)} =
128(125\lambda^{14}-321\lambda^{12}+9525\lambda^{10}+16511\lambda^8-7825\lambda^6-
$$
$$
17995\lambda^4-4001\lambda^2-115)\lambda^4/{15(1-\lambda^2)^{10}}
$$

$$
H^{(7)} =
-256(280\lambda^{18}-45941\lambda^{16}-110888\lambda^{14}+666580\lambda^{12}+
1628568\lambda^{10}-
$$
$$
270014\lambda^8-1470296\lambda^6-524588\lambda^4-37296\lambda^2-245)\lambda^4/{105(1-\lambda^2)^{12}}
$$

$$
H^{(8)} =
64(56\lambda^{22}-169169\lambda^{20}-2072958\lambda^{18}-5222301\lambda^{16}+12116328\lambda^{14}+
$$
$$
35666574\lambda^{12}+3658284\lambda^{10}-29072946\lambda^8-14556080\lambda^6-
$$
$$
1872317\lambda^4-48286\lambda^{2}-49)\lambda^4/{21(1-\lambda^2)^{14}}
$$

$$
H^{(9)} =
2048(37527\lambda^{22}+968829\lambda^{20}+8819501\lambda^{18}+20135431\lambda^{16}-23482698\lambda^{14}-
$$
$$
97554574\lambda^{12}-30319318\lambda^{10}+67137630\lambda^8+46641379\lambda^6+8950625\lambda^4+
$$
$$
495993\lambda^{2}+4683)\lambda^6/{63(1-\lambda^2)^{16}}
$$

$$
H^{(10)} =
-2048(38757\lambda^{26}+1394199\lambda^{24}+31894966\lambda^{22}+243826482\lambda^{20}+
$$
$$
571835031\lambda^{18}-326987427\lambda^{16}-2068579420\lambda^{14}-1054659252\lambda^{12}+
$$
$$
1173787011\lambda^{10}+1120170657\lambda^8+296483526\lambda^6+26886370\lambda^4+
$$
$$
684129\lambda^{2}+2187)\lambda^6/ {45(1-\lambda^2)^{18}}
$$

$$
H^{(11)} =
8192(98142\lambda^{30}-1899975\lambda^{28}+92425520\lambda^{26}+3095961215\lambda^{24}+
$$
$$
25070557898\lambda^{22}+59810870313\lambda^{20}-11635283900\lambda^{18}-173686662185\lambda^{16}-
$$
$$
120533821070\lambda^{14}+74948247123\lambda^{12}+102982107048\lambda^{10}+35567469125\lambda^8+
$$
$$
4673872550\lambda^6+217466315\lambda^4+2569380\lambda^{2}+2277)\lambda^6/{495(1-\lambda^2)^{20}}
$$
\newpage

\begin{figure}
\centerline{ \psfig{figure=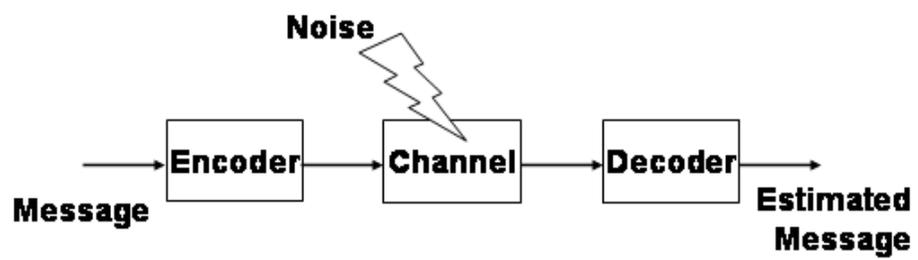,height=5cm} } \caption{
Schematic drawing of message transmission through a noisy channel.
\label{Channel_fig}} \vspace{30cm}
\end{figure}

\vspace{30cm}

\begin{figure}
\centerline{ \psfig{figure=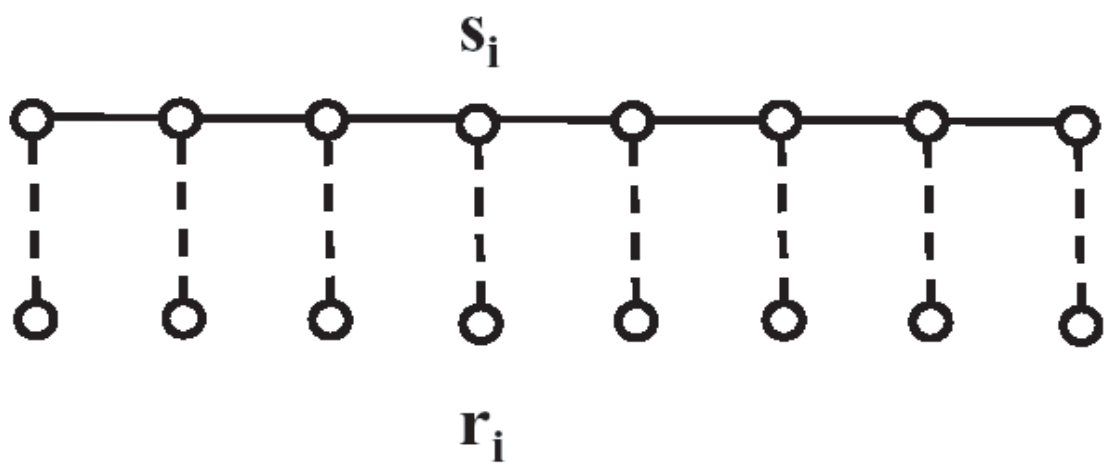,height=6cm} }
\caption{An Ising model in a random field. The solid lines
represent interactions of strength $\mathbf{J}$ between
neighboring spins $\mathbf{S_i}$ while the dashed lines represent
local fields $\mathbf{K r_i}$ acting on the spin $\mathbf{S_i}$.
\label{IsingRandField_fig}} \vspace{30cm}
\end{figure}

\begin{figure}
\hspace*{-0.9cm} {\psfig{figure=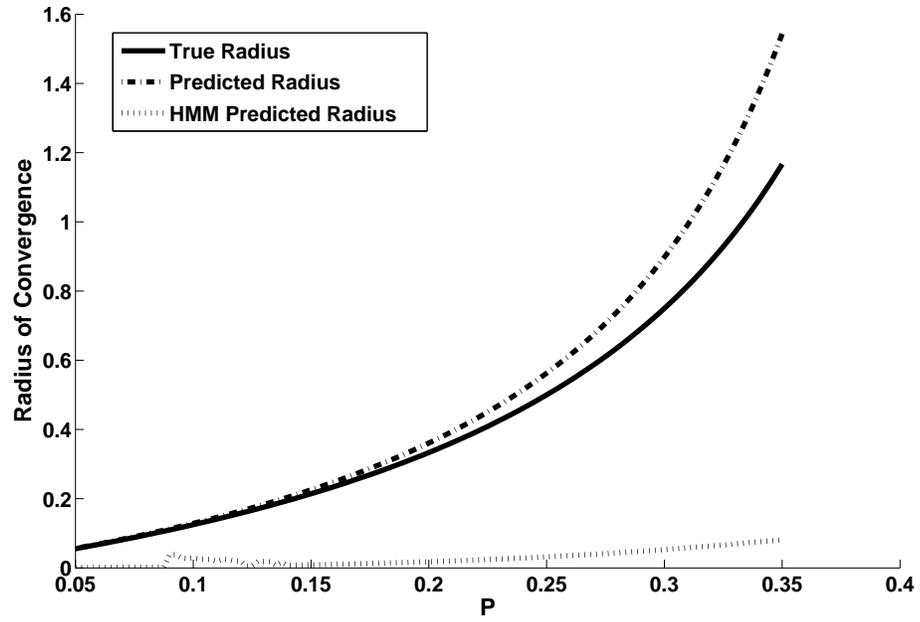,height=10cm}
} \caption{Radius of convergence for the i.i.d. model (estimated
and exact, see text), and {\it HMP} (estimated) for $0.05 \leq p
\leq 0.35$. \label{ConvergenceRadius_fig}} \vspace{30cm}
\end{figure}

\begin{figure}
\hspace*{-1.2cm}
{\psfig{figure=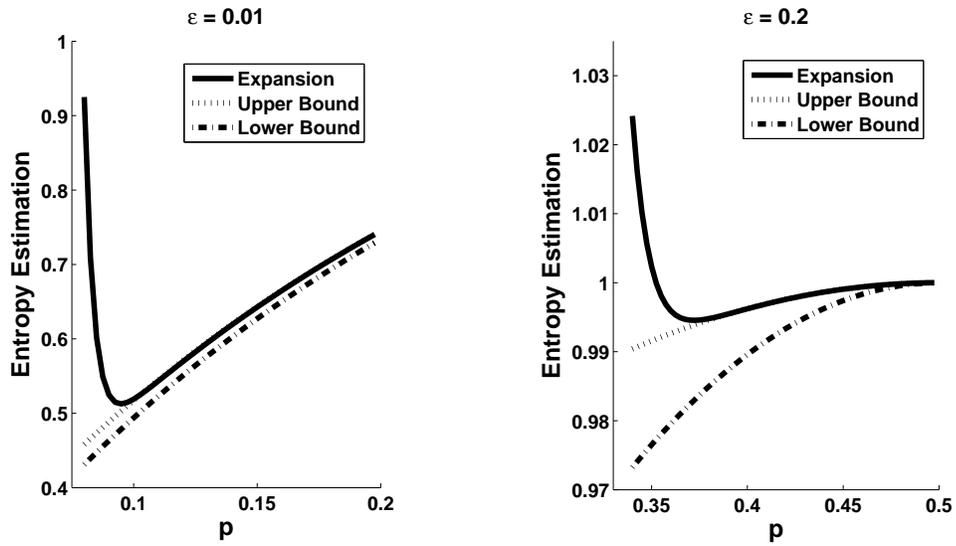,height=10cm} }
\caption{Approximation using the first eleven orders in the
expansion, for $\eps=0.01$ (left) and $\eps=0.2$ (right), for
various values of $p$. For comparison, upper and lower bounds
(using $N=2$ from \cite{Cover}) are displayed. For each $\eps$
there is some critical $p$ below which the series diverges and the
approximation is poor. For larger $p$ the approximation becomes
better. \label{OrdersApproximationTwoEps_fig}} \vspace{30cm}
\end{figure}

%

\end{document}